# Ultimately Thin Metasurface Wave Plates

David Keene[1], Matthew LePain[1], Maxim Durach[1]

[1]*Department of Physics, Georgia Southern University, Statesboro, GA 30460-8031, USA*
*corresponding author: mdurach@georgiasouthern.edu*

Abstract: Optical properties of a metasurface which can be considered a monolayer of two classical uniaxial metamaterials - parallel-plate and nanorod arrays – are investigated. It is shown that such metasurface acts as an ultimately thin sub-50 nm wave plate. This is achieved via an interplay of epsilon-near-zero and epsilon-near-pole behavior along different axes in the plane of the metasurface allowing for extremely rapid phase difference accumulation in very thin metasurface layers. These effects are shown to not be disrupted by non-locality and can be applied to the design of ultrathin wave plates, Pancharatnam-Berry phase optical elements and plasmon-carrying optical torque wrench devices.

Manipulation over polarization of light is one of the major applications of both macroscopic and nanoscopic optical components. The typical working principle of optical wave plates is based on the anisotropic optical response of birefringent media, in which polarization components are separated and accumulate a phase difference upon propagation. The main drawback of this approach is that the phase difference required to apply the needed polarization transformations is acquired over large distances due to low refraction index contrast, making the resulting wave-plate devices bulky [1]. The advance of metasurface photonics happening at the moment has already brought paradigm-changing ideas in the field of optics [2-4]. In particular, new ways to achieve polarization conversion were proposed, including metasurface cavities which allow for reduction of wave plate thickness to a fraction of a micron [5-9]. Several single metasurface structures were proposed as well, providing further reduction in size [10-13]. Nevertheless, the metasurface wave plates proposed to-date feature complex textures and require challenging fabrication. Two classical examples of metamaterials are parallel-plate and nanorod arrays and are known to exhibit strong anisotropy. Recently, it was experimentally shown that the anisotropic properties of epsilon-near-zero (ENZ) nanowire metamaterials lead to polarization rotation in 350-nm-thick metasurfaces based on Au nanorods [14]. The identical structure has been shown to exhibit effective Fabry-Perot (FP) resonances, whose frequencies are governed by effective medium parameters [15].

In this paper we consider a metal nanowire grid metasurface, which can be considered as a monolayer of both parallel-plate array and nanowire array metamaterial (see Fig 1(a)). Two-dimensional wire grid arrays are known as the most commonly used and simplest types of *polarizers*. Here we demonstrate that, surprisingly, such metasurfaces with thickness of just 30-50 nm, despite their simple geometry and nanoscale dimensions, can in fact serve as an efficient *half- and quarter-wave plate* in the near-ir and visible range. This is achieved via interplay of simultaneous ENZ and epsilon-near-pole (ENP) response of these metasurfaces along different axes, allowing for rapid phase difference accumulation. Thicker 100-200 nm structures support a plethora of effective FP resonances for ordinary and extraordinary waves and provide a wide range of opportunities for polarization conversion. In particular, considering wire grid arrays proposed here with inhomogeneous lateral distribution of parameters such as metasurface thickness, metal fraction etc. may lead to a new generation of ultrathin polarization gratings, q-plates and other Pancharatnam-Berry phase optical elements [16-20], which currently attract a lot of attention in the photonics community. Interest in applying torque to nanoscale structures is emerging in such fields as biochemistry [21, 22]. In particular, quartz-based optical torque wrench devices are used [21]. Considering the plasmonic benefits inherent in noble metal nanostructures it would be extremely useful to be able to rotate them optically. The high mass density and the lack of natural anisotropy of noble metals complicate such rotations in comparison to quartz. The ultrathin highly anisotropic metal metasurfaces proposed in this paper will bring such opto-mechanical manipulations into the realm of practicality.

Consider first a parallel-plate metal-dielectric array with period $d = d_m + d_d$, where $d_m = fd$ and $d_d = (1-f)d$ are the thicknesses of the metal and dielectric layers (Fig. 1 (a)). The dielectric permittivities are $\varepsilon_m$ and $\varepsilon_d = n_d^2$ for the metal and dielectric and $f = d_m/d$ is the volumetric metal fraction. The dispersion of the photonic states is described by the Kronig-Penney equation. The photonic states with zero Bloch wave vector $k_x = 0$ can be separated into two symmetry classes for each polarization with dispersions $k_z(\omega)$ given by [23, 24]

$$p_m \tan\left(\frac{\alpha_m d_m}{2}\right) + p_d \tan\left(\frac{\alpha_d d_d}{2}\right) = 0, \quad p_d \tan\left(\frac{\alpha_m d_m}{2}\right) + p_m \tan\left(\frac{\alpha_d d_d}{2}\right) = 0 \quad (1)$$

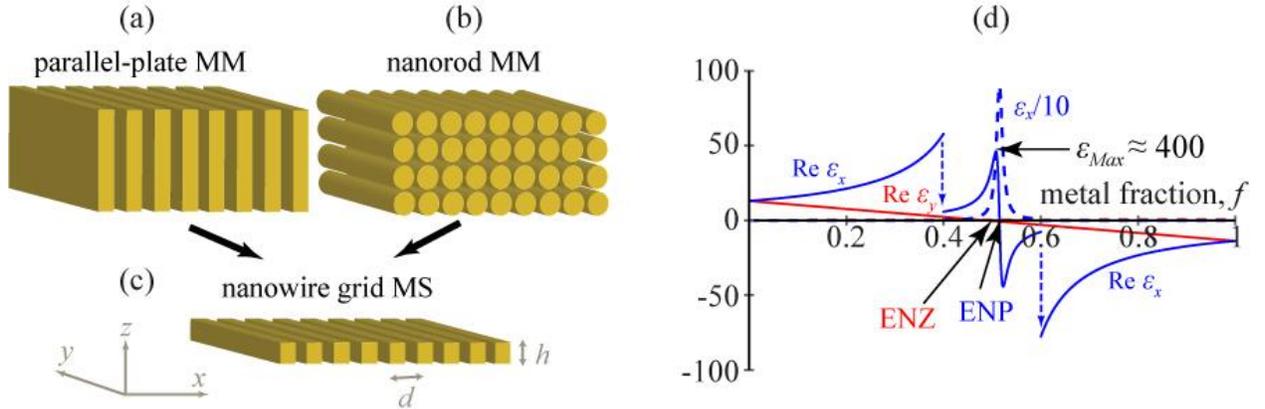

Fig. 1: Schematic of the structure and its typical dielectric response. (a), (b), (c) Correspondence between the parallel-plate and the nanorod metamaterials and the monolayer nanowire grid metasurface considered in this paper. (d) Dielectric permittivities $\varepsilon_x$ (blue for real and dashed blue for imaginary parts) and $\varepsilon_y$ (red) as functions of metal fraction at $\hbar\omega = 2.2\ eV$ for silver/GaAs metasurface, featuring ENP and ENZ.

where $\alpha_{m,d} = \sqrt{k_0^2 \varepsilon_{m,d} - k_z^2}$, $p_{m,d}^{TM} = \alpha_{m,d}/(k_0 \varepsilon_{m,d})$ and $p_{m,d}^{TE} = -\alpha_{m,d}/k_0$ for TM and TE polarizations, with the free space wave vector $k_0 = 2\pi/\lambda_0$.

The dispersion Eqs. (1) have an infinite number of solutions. In a subwavelength array $k_0 d \ll 1$, only modes with $\alpha_{m,d} d_{m,d} \ll 1$ efficiently couple to external radiation incident normally upon the array. In this case, the tangent functions in Eqs. (1) can be substituted by their arguments. Only the first of Eqs. (1) leads to reasonable dispersion equations in this limit, which are given by $k_z^2 = k_0^2 \varepsilon_x$ for TM fields (electric field along *x*-axis in Fig. 1 (a), (c)) and $k_z^2 = k_0^2 \varepsilon_y$ for TE fields (electric field along *y*-axis). The effective dielectric permittivities are given by

$$\varepsilon_x^{-1} = \varepsilon_m^{-1} f + \varepsilon_d^{-1}(1-f), \qquad \varepsilon_y = \varepsilon_m f + \varepsilon_d(1-f). \tag{2}$$

Hence in the subwavelength case the array forms a metamaterial, with 3 bright modes, whose fields follow the effective medium approximation and all the other photonic states are dark plasmonic modes. Using Maxwell-Garnett approximation, the effective dielectric permittivities for the nanorod metamaterial (Fig. 1 (b)) can be derived giving the same expression as in Eq. (2) for $\varepsilon_y$ and a qualitatively similar expression for $\varepsilon_x$, featuring ENP transition [25]. Below we use Eqs. (2) for our effective medium calculations and model given in Fig. 1 (a) and (c) for exact calculations, but the same ideas apply to the nanorod metamaterials or metasurfaces as well. Note that the structures we consider here are different in principle from the structures considered in [14, 15], where the optic axes were perpendicular to the plane of the metasurfaces and the dielectric response in the plane of the metasurface was isotropic, which required large angles of incidence to achieve polarization manipulation. In this paper we consider structures with optic axes in the plane of the metasurfaces perpendicular to the incidence direction, and anisotropy of the dielectric properties *in the plane* of the metasurface is crucial for the effects we predict.

With this in mind, we first model the monolayer metasurface (Fig. 1(c)) as a layer of metamaterial with anisotropic dielectric response. We use the anisotropic characteristic matrix method, described in our recent paper, where we introduced a concept of hyperbolic resonances in metasurface cavities [9]. Here we consider a single metasurface. The metasurface responds to TM polarized normal incidence with incidence plane oriented along the *x* axis ($\phi = 0°$) as an isotropic material with dielectric permittivity $\varepsilon_x$. In the case of $\phi = 90°$ the optical response is fully determined by $\varepsilon_y$. According to the Polarization Rotation Equations [9] the reflection and transmission coefficients for TM and TE polarizations, $r_p^\phi, t_p^\phi$ and $r_s^\phi, t_s^\phi$, at arbitrary $\phi$ can be determined in terms of that for $\phi = 0°$ and $\phi = 90°$ as

$$\begin{aligned} r_p^\phi &= r_p^{0°} \cos^2\phi + r_p^{90°} \sin^2\phi, & t_p^\phi &= t_p^{0°} \cos^2\phi + t_p^{90°} \sin^2\phi, \\ r_s^\phi &= \left(r_p^{0°} - r_p^{90°}\right) \sin\phi \cos\phi, & t_s^\phi &= \left(t_p^{90°} - t_p^{0°}\right) \sin\phi \cos\phi. \end{aligned} \tag{3}$$

Note that at normal incidence the dielectric properties along the *z* axis ($\varepsilon_z = \varepsilon_y$ for parallel plates and $\varepsilon_z = \varepsilon_x$ for nanorods) do not affect the response of the metasurface, which removes the principal difference between the nanorods and parallel-plates.

In a conventional wire grid polarizer metal should be considered a perfect conductor $\varepsilon_m \to i\infty$, resulting in $\varepsilon_y \to i\infty$ and $t_p^{90°} = 0$; then Eqs. (3) are reduced to equations describing a polarizer. The optical properties of plasmonic metals in the visible frequency range are strikingly different from those of perfect conductors and are determined by finite permittivities $\varepsilon_m$ with negative real part Re $\varepsilon_m < 0$. Therefore the nanowire grid metasurface proposed in this paper is not a polarizer. We demonstrate that it can behave as quarter- and half-wave plates.

Consider a purely dielectric slab with thickness $h$ and metal fraction $f = 0$. Typically, such a slab is not transparent with the exception of the Fabry-Perot (FP) resonance frequencies, which are the same for any polarization at normal incidence. Adding metal into the structure as we propose (Fig. 1(a)-(c)) has the opposite effect on modes in different polarizations. For the photonic states with the electric field polarized along $y$ axis ($\phi = 90°$), the effective dielectric permittivity $\varepsilon_y$ is linearly reduced from positive $\varepsilon_d$ to $\varepsilon_m$ with negative real part, passing through the epsilon-near-zero (ENZ) transition at metal fraction (see the red curve in Fig. 1(d))

$$f = f_{ENZ}(\omega) = \frac{\varepsilon_d}{\varepsilon_d - \mathrm{Re}\varepsilon_m}. \tag{4}$$

Correspondingly, for $f < f_{ENZ}$ the effective wavelength of the *ordinary* waves $\lambda_0/\sqrt{\varepsilon_y}$ is increased as metal is added, culminating in a virtually constant field at ENZ.

For the states with extraordinary polarization ($\phi = 0°$), the effective dielectric permittivity $\varepsilon_x$ exhibits a non-monotonous behavior, featuring the epsilon-near-pole (ENP) transition during which the effective medium transforms from dielectric phase into a metal, passing through a region of strong absorption for extraordinary waves. The ENP transition (see the blue curve in Fig. 1(d)) occurs at metal fraction

$$f = f_{ENP}(\omega) = \frac{\mathrm{Re}\varepsilon_m}{\mathrm{Re}\varepsilon_m - \varepsilon_d} = 1 - f_{ENZ}, \tag{5}$$

For $f < f_{ENP}$ the dielectric permittivity $\varepsilon_x$ increases with the increase of $f$ reaching the maximum real value of $\varepsilon_{Max}$ (Fig. 1(c)). This leads to a decrease in the effective wavelength $\lambda_0/\sqrt{\varepsilon_x}$ of the *extraordinary* waves. Note that the values of $\varepsilon_{Max}$ can be very high, for example, $\mathrm{Re}\,\varepsilon_x \approx 400$ in Fig. 1(d). This is an artifact of the effective medium approximation, since for such values of $k_z/k_0 = \sqrt{\varepsilon_{Max}}$ the condition $\alpha_{m,d}d_{m,d} \ll 1$ is not satisfied. Nevertheless, as we show below by semi-analytical solution of Maxwell equations for the structure in Fig. 1 (a) or (c), this leads to moderate quantitative deviation with no qualitative modification of our results.

The FP resonance energies $\hbar\omega_o$ ($\phi = 90°$) for ordinary (o-FP$_n$) and $\hbar\omega_e$ ($\phi = 0°$) extraordinary modes (e-FP$_n$) depend on metal fraction $f$ and thickness of the metasurface $h$ as

$$\hbar\omega_{on} = \frac{\hbar c}{\sqrt{\varepsilon_y(\omega_{on},f)}}\left(\frac{\pi n}{h}\right), \qquad \hbar\omega_{en} = \frac{\hbar c}{\sqrt{\varepsilon_x(\omega_{en},f)}}\left(\frac{\pi n}{h}\right), \quad n \text{ is an integer} \tag{6}$$

The conditions of the FP resonances given by Eq. (6) can be resolved for the metal fraction as

$$f_{on} = f_{ENZ}(1 - n^2\xi^2), \qquad f_{en} = f_{ENP}\left(1 - \frac{1}{n^2\xi^2}\right), \qquad \xi = \frac{\lambda_0}{2n_d h} \tag{7}$$

This means that o-FP$_n$ approach the ENZ transition for small $n$, with $n = 0$ corresponding to ENZ exactly, while e-FP$_n$ approach the ENP in the limit $n \to \infty$. We confirm this by calculations in the effective medium approximation first. Using a $4 \times 4$ characteristic matrix approach [9], we calculate the reflectivity of the metasurface, which is shown in Fig. 2. Note that for calculations in this paper we use the dielectric permittivity of silver [26] and a dielectric with refractive index $n_d = \sqrt{\varepsilon_d} = 3.6$, which corresponds to GaAs.

For a pure dielectric $f = 0$ the reflection spectrum of the metasurface is composed of FP reflectivity dips separated by broad bands of strong reflection for both $\phi = 0°$ (Fig. 2 (a)) and at $\phi = 90°$ (Fig. 2 (b)). As metal is inserted $e\text{-FP}_n$ at $\phi = 0°$ and $o\text{-FP}_n$ at $\phi = 90°$ split and closely follow Eqs. (6)-(7) (outlined by the dashed lines in Fig. 2), confirming our predictions.

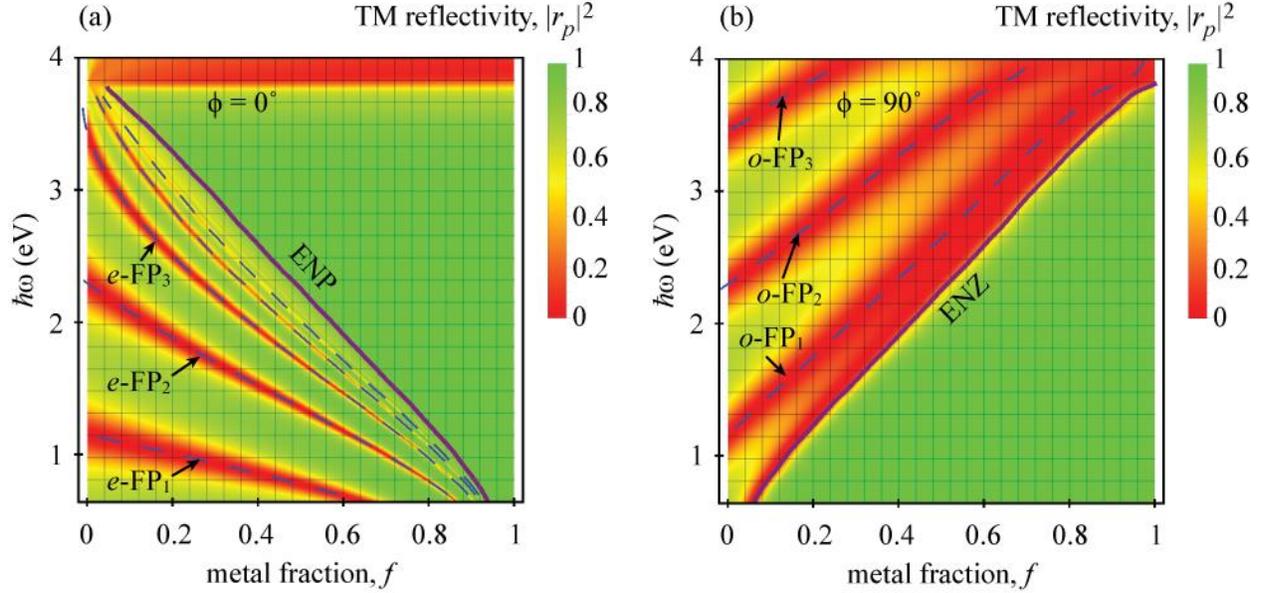

Fig. 2: Reflectivity of a 150-nm-thick metasurface as a function of the metal fraction $f$ and energy of the incident radiation $\hbar\omega$ at (a) $\phi = 0°$ and (b) $\phi = 90°$ calculated using the effective medium approximation.

The energies of the FP resonances depend strongly on $f$, such that the $n$th extraordinary $e\text{-FP}_n$ mode may intersect $m$th ordinary $o\text{-FP}_m$ resonance if $n > m$ at energies which we denote as $E_{hw,mn}$, where "$hw$" stands for "half-wave plate" as will be explained below. Considering Eq. (3) for $r_p^\phi$ these intersections should explicitly appear at $\phi = 45°$. We plot the TM reflectivity $\left|r_p^{45°}\right|^2$ in Fig. 3 for metasurfaces with different thicknesses.

In Fig. 3(a) we plot reflectivity for a very thin metasurface with $h = 30$ nm. Since in this case $\xi \gg 1$ only one extraordinary $e\text{-FP}_1$ resonance with $f \approx f_{ENP}$ is visible. Higher order resonances are positioned closer to ENP and are extremely faint due to strong absorption in the effective medium at ENP. The only ordinary wave resonance present has $n = 0$ and corresponds to ENZ. The $e\text{-FP}_1$ and the ENZ intersect at $\hbar\omega = E_{hw,01}$ as indicated by the black dot. We show reflectivity for thicker metasurfaces $h = 100$ nm and 150 nm in Fig. 3 (b)-(c). As one can see the $e\text{-FP}_1$ shifts to lower energies and higher order extraordinary resonances become strong. Additionally ordinary resonances appear and intersect with higher order extraordinary resonances at $\hbar\omega = E_{hw,mn}$. The metal fractions $f_{mn}$ at which these intersections occur can be found from the condition $f_{om} = f_{en}$ in the following form

$$f_{mn} = \frac{1 - n^2\xi^2}{1 - n^2\xi^2(2 - m^2\xi^2)} \tag{8}$$

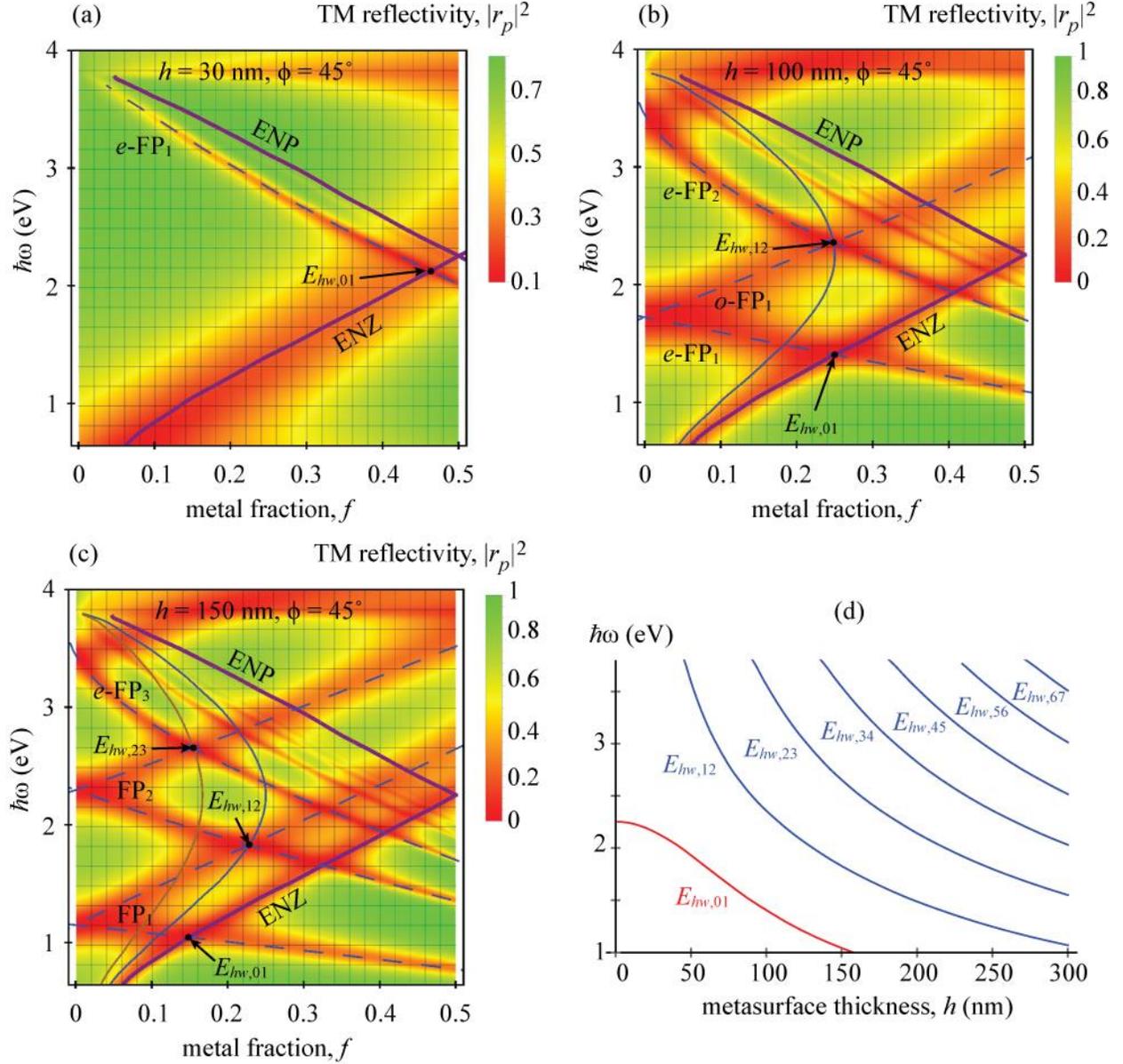

Fig. 3: Reflectivity of metasurfaces at $\phi = 45°$. (a) $h = 30\ nm$; (b) $h = 100\ nm$; (c) $h = 150\ nm$. (d) Dependence of the energies $E_{hw,nn+1}$ at which the metasurface is a half-wave plate on metasurface thickness.

We are mainly interested in intersections between the consecutive FP resonances at $E_{hw,nn+1}$. For very thin metasurfaces $\xi \gg 1$ the intersection between $e$-FP$_1$ and the ENZ occurs at $f_{01} \approx 0.5$ and with increase of the thickness $h$ shifts down in energy along the ENZ line as shown in Figs. 3 (a)-(c). We plot the functions $f_{12}(\omega)$ and $f_{23}(\omega)$ in Fig. 3 (b)-(c) as blue and brown thin solid lines. The intersections $E_{hw,nn+1}$ are marked by black dots and as $h$ is increased their energies are reduced, moving along the $f_{nn+1}(\omega)$ curves. This dependence of $E_{hw,nn+1}$ on metasurface thickness $h$ is shown in Fig. 3(d). For thicker metasurfaces more FP resonances and their intersections appear in the plasmonic frequency range of silver (~1-4 eV).

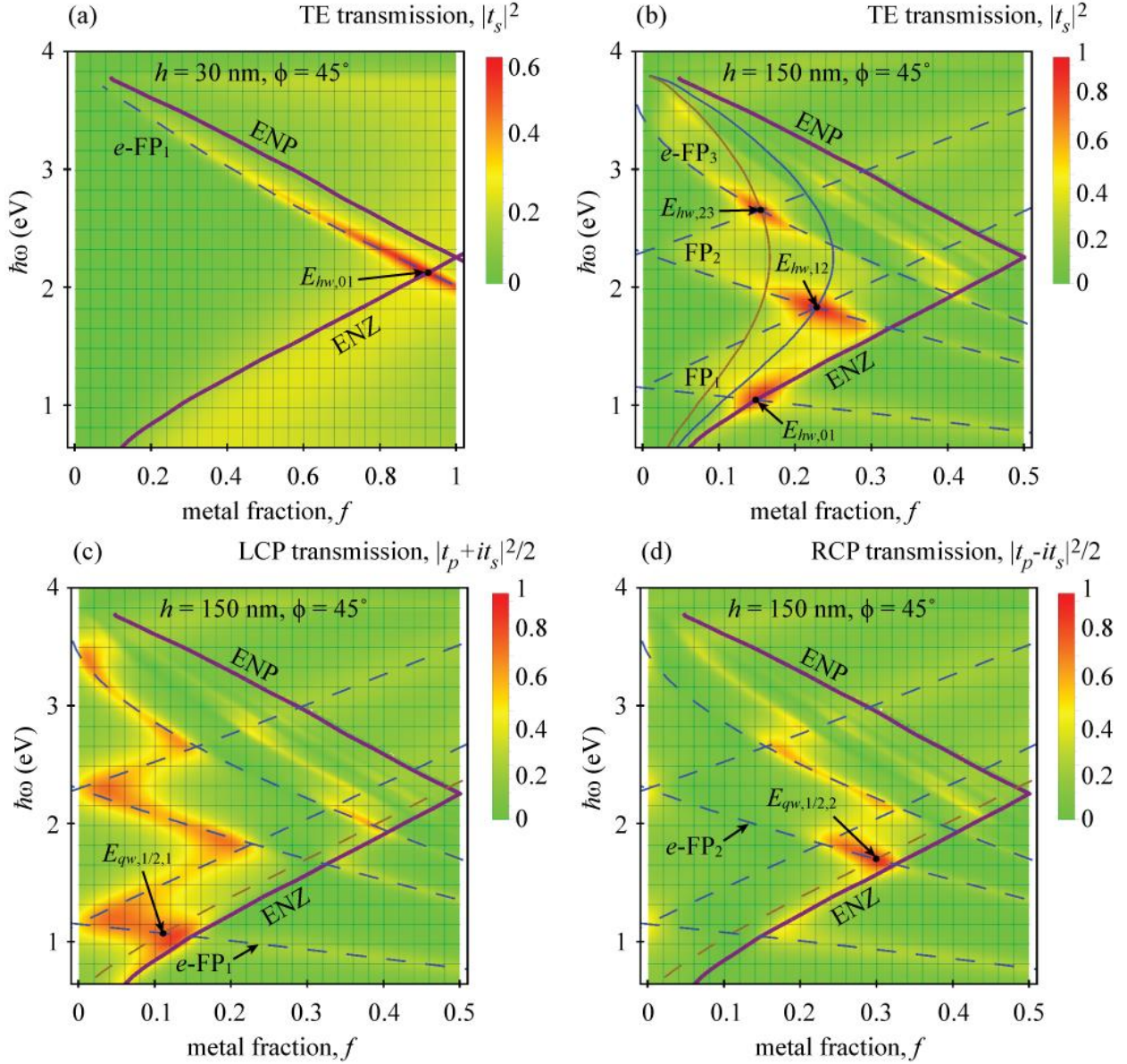

Fig. 4: (a), (b) TE transmission for $h = 30\ nm$ and $h = 150\ nm$ in response to TM polarized light. (c), (d) Transmission of left and right circular polarized light in response to TM polarized incidence for metasurface thickness $h = 150\ nm$.

The values of the transmission coefficients $t_p$ for consecutive FP modes alternate in sign and are close to unity in magnitude. We can conclude from Eqs. (3) that at the intersections between $o$-$FP_n$ and $e$-$FP_{n+1}$ when the incidence plane is at angle $\phi = 45°$, the TM polarized incident radiation undergoes a 90-degree polarization rotation. Indeed, in this case $\left|t_p^{45°}\right|^2 \approx 0$ and $\left|t_s^{45°}\right|^2 \approx 1$. Similarly polarization rotation occurs when $e$-$FP_{n+1}$ intersect ENZ. In Fig. 4(a) we

plot TE transmission $\left|t_s^{45°}\right|^2$ for $h = 30$ nm. Comparing this figure with Fig. 3(a) we confirm the polarization rotation with efficiency $\left|t_s^{45°}\right|^2 = 0.6$ at intersection of $e$-FP$_1$ and the ENZ at a remarkably thin metasurface. Note that loss of about 40% of the intensity of incident radiation is due to high losses near the ENP transition (as seen in Fig. 1 (d)), since point $(f_{01}, E_{hw,01})$ is located next to ENZ-ENP intersection (see Fig. 4 (a)).

The nature of polarization rotation in the ENZ-ENP structure (Figs. 3(a) and 4(a)) which we propose here is easy to understand. The electric field of the ENZ ordinary wave directed along the $y$-axis (see schematics in Fig. 1 (a)-(c)) exhibits no phase change, while the extraordinary fields directed along the $x$-axis oscillate with wavelength $\lambda_x = \lambda_0/\sqrt{\varepsilon_{Max}} = 2h$, providing the required phase difference for the wave-plate effect in an extremely thin metasurface. We call this structure an *ultrathin ENZ-ENP wave plate*.

A larger thickness $h = 150$ nm metasurface (Fig. 4 (b)) has intersections at $\hbar\omega = E_{hw,nn+1}$ that lead to peaks with $\left|t_s^{45°}\right|^2 > 0.95$ in TE transmission. Hence we observe a half-wave plate behavior in a 150-nm-thick metasurface with close-to-100% efficiency. Further investigating this structure, we can find conditions in which the difference in phase shifts for ordinary and extraordinary waves lead to quarter-wave-plate behavior, converting linearly polarized incident radiation to circularly polarized transmission. This effect is strongest at intersections of $\hbar\omega_{o,1/2}$ and $\hbar\omega_{en}$ as can be seen in Fig. 4 (c)-(d). For $h = 150$ nm the efficient close-to-90% conversion into left-circular polarized (LCP) radiation happens at $E_{qw,1/2,1}$, while 90% conversion into right-circular polarized (RCP) occurs at $E_{qw,1/2,2}$. This high-efficiency conversion from linear to circular polarization indicates that the metasurfaces proposed here can acquire torques from the incident radiation with power $P_{inc} = $ mW of about $\tau \approx P_{inc}/\omega$ in the $10^3$ pN·nm range, similar to the optical torque wrench structures previously proposed [21]. At the same time these metasurfaces have moments of inertia of comparable magnitudes, despite being composed of heavy Au atoms due to reduced dimensionality. This introduces the possibility of high-rpm rotation of plasmonic metal nanostructures.

We confirm our results obtained using the effective medium approximation by solving Maxwell's equations semi-analytically for the nanostructure shown in Figs. 1 (a) and (c) using our recently developed numerical method [24]. To do so we first find the solution to the Kronig-Penney equation, which for normal incidence is given by Eqs. (1). As was noted before, if $\alpha_{m,d}d_{m,d} \ll 1$, these equations lead to the effective medium dispersion relations. The effective permittivity may become large at ENP, which leads to $k_z \approx \alpha_{m,d} \gg 1/d_m$, violating the applicability conditions for the effective medium approximation. We compare the exact solution of Eq. (1) for the bright extraordinary modes in the structure with period $d = 50$ nm with the effective medium dispersion, obtained using $\varepsilon_x$ from Eq. (2) (see Fig. 1(d)). We are interested in deviations of the mode propagation wave vector $k_z$, since $k_z = n\pi/h$ is the condition for the extraordinary FP resonances.

We provide the three most extreme examples of such deviations for the two resonances: the $e$-FP$_1$ mode at $h = 30$ nm which requires $k_z = \pi/(30$ nm$)$ (shown as the right red dashed line in panels of Fig. 5) and the $e$-FP$_3$ mode at $h = 150$ nm requiring $k_z = 3\pi/(150$ nm$)$ (the left red dashed line in Fig. 5). The blue curves in Fig. 5 represent the exact solution of Eq. (1), while the green curves are the effective medium approximation. In Fig. 5(a) for $f = 0.3$ the exact solution intersects the $n\pi/h$ lines at lower frequencies than the effective medium approximation, as indicated by the blue and green arrows. The real part of the exact solution does not intersect the $k_z = \pi/(30 \text{ nm})$ line, but comes close to it at $\hbar\omega = 2.33$ eV, so the resonance at this frequency should be expected.

In Fig. 5(b) for $f = 0.4$ we see that the exact dispersion coincides with the effective medium dispersion at $k_z = 3\pi/(150 \text{ nm})$. Meanwhile, for the exact dispersion at Re $k_z = \pi/(30 \text{ nm})$ two resonances exist within close proximity at $\hbar\omega = 2.1$ eV and $\hbar\omega = 2.2$ eV due to back-bending, instead of a single frequency for the effective medium at $\hbar\omega = 2.33$ eV. As can be seen from Fig. 5(c) the high-frequency root from the back-bending pair with Re $k_z = \pi/(30 \text{ nm})$ has much higher damping as it is overwhelmed by large Im $k_z$ (represented by the dashed blue line), leaving only the low-frequency root as a resonance. It is important to note that this is always the case for the effective medium approximation regardless of $f$. For $f = 0.5$ the exact solution intersects the $n\pi/h$ lines at higher frequencies than the effective medium dispersion.

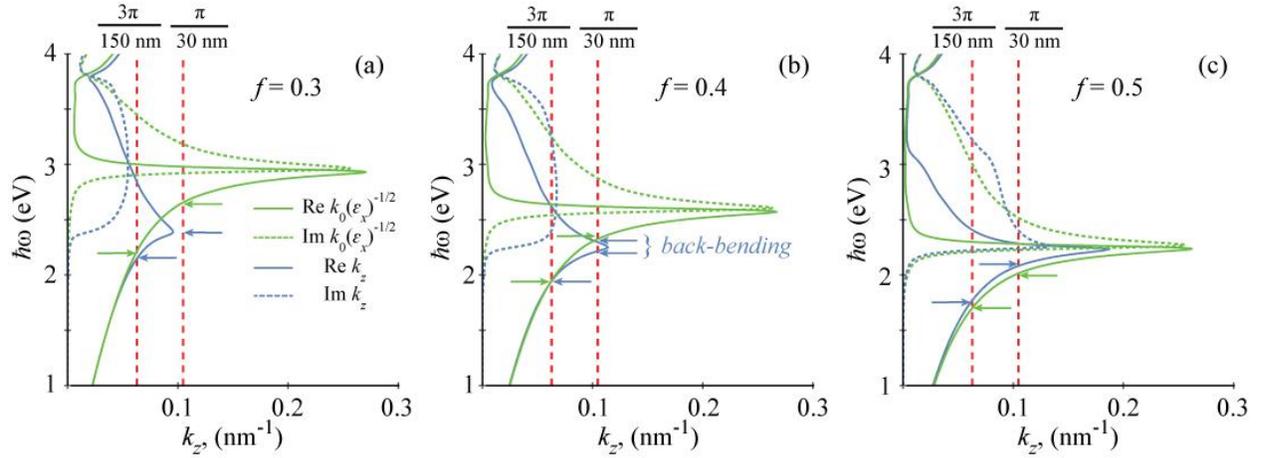

Fig. 5: Comparison of the exact solution $\omega(k_z)$ of Eq. (1) for period $d = 50\,nm$ and the effective medium approximation (the inverse function of $k_0\sqrt{\varepsilon_x(\omega)}$) with (a) $f = 0.3$; (b) $f = 0.4$; (c) $f = 0.5$.

This analysis directly translates into the optical response of the metasurface in the full solution of Maxwell's equations according to semi-classical method of Ref. [24] illustrated in Fig. 6. Panels (a) and (b) show the reflectivity of the structure at $h = 30$ nm and $h = 150$ nm respectively, which correspond directly to the effective medium calculations shown in Fig 3 (a) and (c). Similarly, panels (c) and (d) in Fig. 6 show TE transmittance in direct comparison to Fig. 4(a) and (b). The ramifications of the disparity between $k_z$ calculated exactly and in the effective medium approximation near the ENP as outlined in Fig. 5 make themselves apparent in Fig.6 by the higher degree of curvature and back-bending in the energy as a function of metal fraction $f$ for the $e$-FP$_1$ at $h = 30\,nm$ and $e$-FP$_3$ at $h = 150\,nm$ (Fig. 6 (a) and (b) respectively). This results in minor shifts in the parameters and insignificant changes in the magnitude for the half-wave (Fig. 6 (c) and (d)) and quarter-wave plate behavior (not shown) as manifested by the TE transmission at $\hbar\omega = E_{hw,nn+1}$.

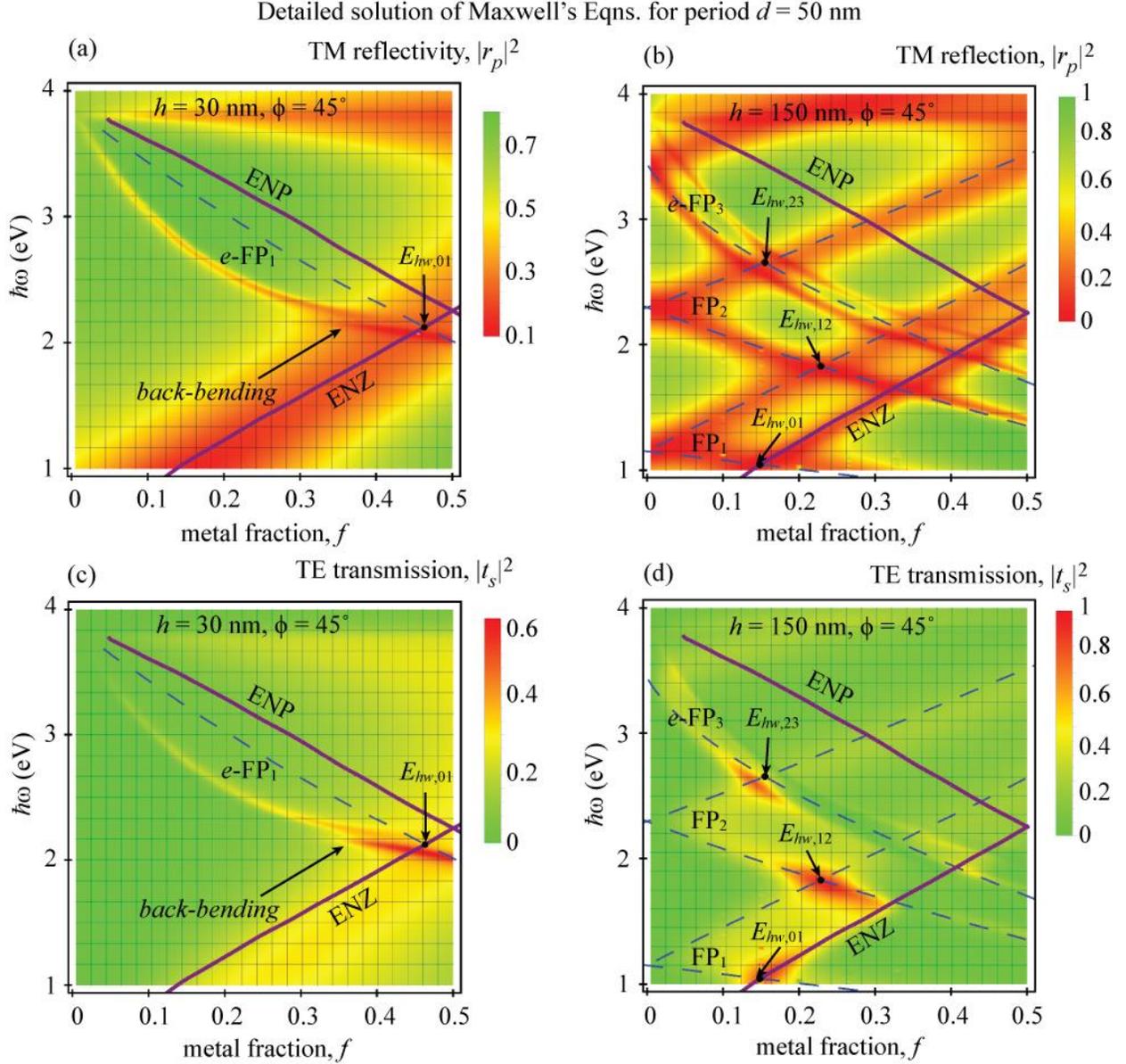

Fig. 6: Optical properties on the metasurface shown in (Fig. 1(c)) according to the semi-analytical solution of Maxwell's equations. (a), (b) Reflectivity of the metasurface at $\phi = 45°$ for $h = 30\ nm$ and $h = 150\ nm$; (c), (d) TE transmission of the same metasurfaces.

All the calculations shown before this point are for strictly normal incidence for the sake of simplicity as that is the most natural orientation for wave plates. We would like to demonstrate now that there is a range of incidence angles in which those properties persist. In Fig. 7 we show the 2-dimensional dispersion of the FP modes for $h = 150$ nm structure. For $f = 0.1$ (Fig. 7(a)) the two FP resonances shown (defined by $|r_p|^2 < 0.01$) do not intersect and feature highly anisotropic dispersion oriented along the respective axes. When these resonances intersect (Fig. 7(b)) they interfere to produce increased polarization rotation with 70% conversion efficiency in the $\pm 10°$ range of angles around $\phi = 45°, 135°, 225°, 315°$ (see the red clover-leaf contour). In the other directions the structure is transmitting without polarization rotation as shown by the green contour. Note that all of the resonances exhibit an anisotropics parabolic dispersion.

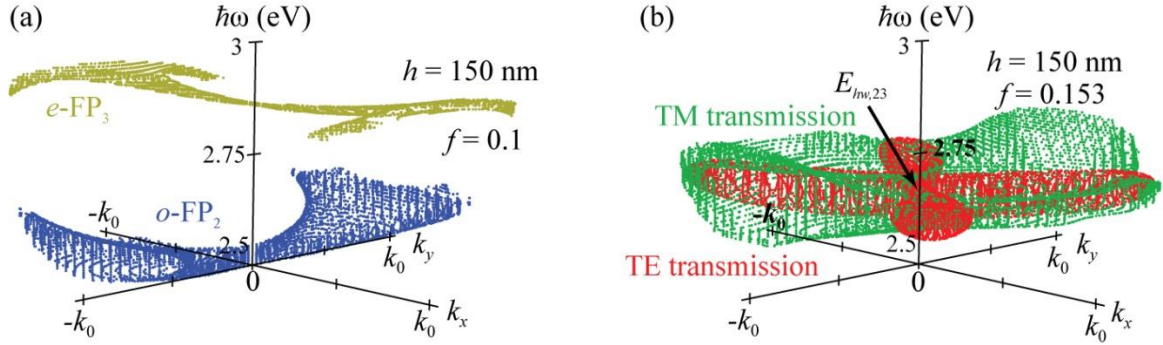

Fig. 7: Dispersion of the *e-FP*$_3$ and *o-FP*$_2$ resonances in $h = 150\ nm$ structure (a) off and (b) at the intersection $\hbar\omega = E_{hw,23}$.

In conclusion, we propose a nanowire grid metasurface, which can serve as an ultrathin ENZ-ENP wave plate as is shown in the effective medium approximation and confirmed by semi-analytical solution of Maxwell's equations. Polarization manipulation accomplished on sub-50 nm distances predicted here opens new possibilities for the fields of photonics and nanoscopic opto-mechanics.

Acknowledgements: This study was supported by funds from the Office of the Vice President for Research & Economic Development and the Jack N. Averitt College of Graduate Studies at Georgia Southern University. D. K. and M. L. are grateful for the support provided by the student research grant from College Office of Undergraduate Research (COUR) at Georgia Southern University. M. L. is grateful for the support of the Blue Waters Internship program. M. D. is thankful to Dr. Natalia Noginova for useful discussions.

References:

[1]  M. Born, E. Wolf, *Principles of Optics* (Cambridge University Press, 1999).
[2]  A. V. Kildishev, A. Boltasseva, V. M. Shalaev, *Science* 339 (6125), (2013).
[3]  N. Yu, F. Capasso, Nature Materials **13**(2): 139-150 (2014)
[4]  Y. Zhao, X.-X. Liu, A. Alù, Journal of Optics 16(12): 123001, (2014)
[5]  J. Hao, Q. Ren, Z. An, X. Huang, Z. Chen, M. Qiu, L. Zhou, Physical Review A 80(2): 023807 (2009).
[6]  A. Pors, M. G. Nielsen, S. I. Bozhevolnyi, Optics Letters 38(4): 513-515 (2013)
[7]  A. Pors, S. I. Bozhevolnyi, Optics Express 21(3): 2942-2952 (2013)
[8]  Z. H. Jiang, L. Lin, D. Ma, S. Yun, D. H. Werner, Z. Liu, T. S. Mayer, Scientific Reports 4 (2014).
[9]  D. Keene, M. Durach, Optics Express 23, 18577-18588 (2015)
[10] A. Papakostas, A. Potts, D. M. Bagnall, S. L. Prosvirnin, H. J. Coles and N. I. Zheludev, Physical Review Letters 90(10): 107404 (2003)
[11] A. Drezet, C. Genet and T. W. Ebbesen, Physical Review Letters 101(4): 043902 (2008)
[12] Khoo, E. H., E. P. Li and K. B. Crozier, Optics letters 36(13): 2498-2500 (2011)
[13] Y. Zhao, A. Alù, Physical Review B 84(20): 205428 (2011).

[14] P. Ginzburg, F. J. R. Fortuño, G. A. Wurtz, W. Dickson, A. Murphy, F. Morgan, R. J. Pollard, I. Iorsh, A. Atrashchenko, P. A. Belov, Y. S. Kivshar, A. Nevet, G. Ankonina, M. Orenstein and A. V. Zayats, Optics Express 21(12): 14907-14917 (2013)

[15] Slobozhanyuk, A. P., P. Ginzburg, D. A. Powell, I. Iorsh, A. S. Shalin, P. Segovia, A. V. Krasavin, G. A. Wurtz, V. A. Podolskiy, P. A. Belov, A. V. Zayats, Physical Review B 92(19): 195127 (2015)

[16] Franco Gori, Optics Letters, 24, 584-586 (1999)

[17] Uriel Levy, Hyo-Chang Kim, Chia-Ho Tsai, and Yeshaiahu Fainman, Optics Letters 30, 2089-2091 (2005)

[18] G. Zheng, Mühlenbernd, H., Kenney, M., Li, G., Zentgraf, T., S. Zhang, Nature Nanotechnology, 10(4), 308-312 (2015)

[19] L. Marrucci, C. Manzo, D. Paparo, Physical Review Letters, 96(16), 163905 (2006).

[20] Z. E. Bomzon, G. Biener, V. Kleiner, E. Hasman, Optics Letters, 27(13), 1141-1143 (2002)

[21] A. La Porta, M. D. Wang, Physical Review Letters 92(19): 190801 (2004)

[22] F. Pedaci, Z. Huang, M. van Oene, S. Barland, N. H. Dekker, Nature Physics 7(3): 259-264 (2011)

[23] B. Sturman, E. Podivilov, M. Gorkunov, Phys. Rev. B, 77(7):075106 (2008)

[24] M. LePain, M. Durach, accepted to Journal of Computational Science Education, Vol. 7; arXiv:1508.01861 (2015)

[25] J. Elser, R. Wangberg, V. A. Podolskiy, E. E. Narimanov, Applied Physics Letters 89(26): 261102 (2006).

[26] P. B. Johnson, R. W. Christy, Physical Review B 6(12): 4370-4379 (1972)